\documentclass[sigconf,10pt,nonacm]{acmart}
\AtBeginDocument{%
	}

\usepackage{graphicx}
\usepackage{xcolor}
\usepackage{xspace}
\usepackage{tikz}

\usepackage[utf8]{inputenc}
\usepackage[T1]{fontenc}

\usepackage{listings}
\lstdefinestyle{customdsl}{
  language=C,
  basicstyle=\ttfamily\scriptsize,
  keywordstyle=\color{blue},
  commentstyle=\color{green},
  numberstyle=\tiny\color{gray},
  stringstyle=\color{red},
  numbers=left,
  numbersep=5pt,
  tabsize=2,
  showstringspaces=false,
  breaklines=true,
  morekeywords={in},
  emph={len, range},
  emphstyle=\color{brown}
}

\usepackage{fontawesome5}
\newboolean{showcomments}
\setboolean{showcomments}{true}
\ifthenelse{\boolean{showcomments}}
{ \newcommand{\mynote}[3]{
    \fbox{\bfseries\sffamily\scriptsize#1}
{\small$\blacktriangleright$\textsf{\emph{\color{#3}{#2}}}$\blacktriangleleft$}}}
{ \newcommand{\mynote}[3]{}}

\newcommand{\subpoint}[1]{\smallskip\noindent\textbf{#1}\xspace}

\title{Towards a DSL for hybrid secure computation}

\author{Romain de Laage}
\affiliation{
	\institution{University of Neuchâtel}
	\country{Switzerland}}
\email{romain.delaage@unine.ch}
\orcid{0009-0009-3476-1688}

\date{\today}

\begin{abstract}
Fully homomorphic encryption (FHE) and trusted execution environments (TEE) are two approaches to provide confidentiality during data processing.
Each approach has its own strengths and weaknesses.
In certain scenarios, computations can be carried out in a hybrid environment, using both FHE and TEE.
However, processing data in such hybrid settings presents challenges, as it requires to adapt and rewrite the algorithms for the chosen technique.
We propose a domain-specific language (DSL) for secure computation that allows to express the computations to perform and execute them using a backend that leverages either FHE or TEE, depending on what is available.
\end{abstract}

\newcommand{\copyrighttext}{  \scriptsize \textcopyright 2025 ACM.
	Personal use of this material is permitted.
	Permission from ACM must be obtained for all other uses,
	in any current or future media, including reprinting/republishing this
	material for advertising or promotional purposes, creating new collective
	works, for resale or redistribution to servers or
	lists, or reuse of any copyrighted component of this work in other works.
	This is the author's pre-print version of the work. Published in the 19th ACM International Conference on Distributed and Event-Based Systems (DEBS).
}

\begin{document}

\newcommand{\copyrightnotice}{\begin{tikzpicture}[remember picture,overlay]
	\node[anchor=south,yshift=2pt,fill=yellow!20] at (current page.south) {\fbox{\parbox{\dimexpr\textwidth-\fboxsep-\fboxrule\relax}{\copyrighttext}}};
	\end{tikzpicture}
}
\maketitle
\copyrightnotice

\section{Introduction}
\label{sec:intro}

Fully homomorphic encryption (FHE) is a cryptographic approach that enables the evaluation of arbitrary functions $f$ over encrypted data. This capability allows users to outsource computations to cloud providers without exposing sensitive cleartext input, intermediate results, or final outputs. However, this method is resource-intensive, raising significant challenges, particularly in terms of computational efficiency and scalability~\cite{delaage25practicalsecureaggregation}.

Conversely, trusted execution environments (TEE) offer a complementary solution through hardware-based secure enclaves that permit data processing in a privacy preserving manner.

The support of both FHE and TEE presents a promising avenue for enhancing data privacy in hybrid environments, where the choice of the secure computation method may depend on factors such as legacy hardware compatibility or the necessity for zero-trust architectures.

Despite the advantages of both FHE and TEE, the practical implementation of secure computations in hybrid environments is challenging. While running legacy applications within TEE-based enclaves is made easy through LibOS approaches, adapting source code and algorithms for FHE often requires significant modifications. This dual maintenance of codebases for both FHE and TEE can lead to increased complexity and resource overhead.

To address these challenges, we propose a domain-specific language (DSL) designed for hybrid secure computation. 
This DSL provides a common representation of computations, enabling developers to write a single codebase that can be executed across multiple backends, whether utilizing FHE or TEE. 
By streamlining the development process and reducing the need for separate implementations, our approach enhances the efficiency and practicality of secure computations in hybrid environments, ultimately helping to increase the adoption of privacy preserving techniques in real-world applications.

\section{Adversarial model}
\label{sec:adv-model}

We consider a \emph{semi-honest (honest but curious)} adversarial model where the interpretor is guaranteed to follow the protocol specification but may attempt to glean additional information from the metadata and the execution patterns of the program.

The type of data and metadata, such as the size of a vector, is explicitly excluded from the scope. This decision is based on a trade-off between security and usability. Additionally, the user may opt to explicitly designate certain data as clear when necessary for runtime operations. In fact, FHE does not allow for data-dependent branching, meaning that data used for loop conditions must be known, for instance.

We consider the standard TEE adversary model where the privileged OS and other hardware is under complete control of the adversary, with the exception of the CPU. Moreover, we assume the presence of a public-key infrastructure as well as cryptographic primitives that make secure communication channels possible.

\subpoint{Discussion.}
Certain metadata could be protected. For instance, the vector size might be substituted with a combination of a maximum size and padding. However, this approach comes with a cost, as it requires more memory space and results in larger messages.

\section{Architecture}
\label{sec:arch}

We propose a DSL specifically designed for secure hybrid computation. This language brings a clear differentiation between encrypted data and cleartext data, traditional control structures adapted to the constraints of FHE, and consistent behavior across the different backends.

\begin{figure}[t!]
	\centering
	\includegraphics[width=\linewidth]{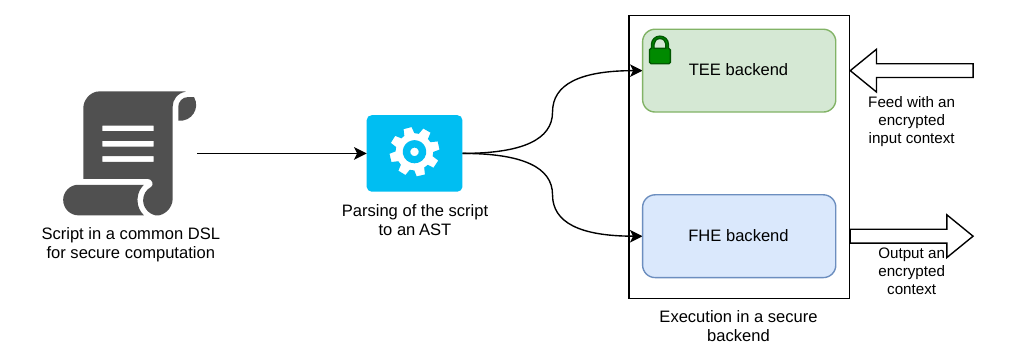}
	\caption{Overview of the DSL}
	\label{fig:dsl-sc-workflow}
\end{figure}

The tool parses the source code into a unified abstract syntax tree (AST), which is then executed by the chosen backend. All backends include an interpreter capable of importing an encrypted context and executing the AST, producing an encrypted output context. This facilitates integration into a data processing workflow. The operation of the DSL is illustrated in \autoref{fig:dsl-sc-workflow}.

The resulting language has some limitations due to the use of FHE for one of the backends. Indeed, if-else statements are treated as expressions, with each branch containing only a single expression. Additionally, encrypted data-dependent branching is not allowed, meaning that loop conditions must operate on vectors of a known size.

\section{Preliminary results}
\label{sec:results}

\autoref{fig:dsl-code} shows the shared code in our DSL to compute the covariance between two arrays of integers (\texttt{xVec} and \texttt{yVec}) provided in the input context of the interpreter. Its output context contains the result in the \texttt{covariance} variable. This code can be run directly on the different backends.

\begin{figure}[t!]
	\begin{lstlisting}[style=customdsl]
    xSum = 0
    for x in xVec {
      xSum = xSum + x
    }
    ySum = 0
    for y in yVec {
      ySum = ySum + y
    }
    xMean = xSum / len(xVec)
    yMean = ySum / len(xVec)
    sum = 0
    for i in range(len(xVec)) {
      sum = sum + (xVec[i] - xMean) * (yVec[i] - yMean)
    }
    covariance = sum / len(xVec)
	\end{lstlisting}
	\caption{Code to get the covariance between two arrays}
	\label{fig:dsl-code}
\end{figure}

\begin{figure}[t!]
	\centering
	\includegraphics[width=\linewidth]{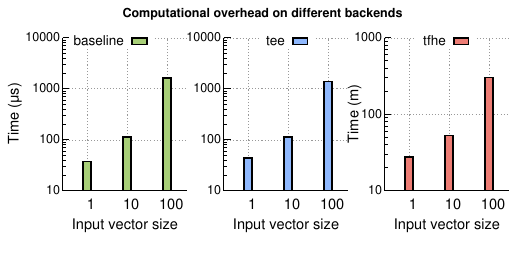}
	\caption{Execution time on different backends}
	\label{fig:plot-dsl}
\end{figure}

\autoref{fig:plot-dsl} shows the execution time of the give algorithm on the different backends, the baseline running without any protection.\footnote{Benchmarks are run on a server with a 16-core Intel(R) Xeon(R) Gold 6326 CPU clocked at 2.90\;GHz, a 24\;MB last-level cache and 64\;GB of DRAM. The server runs Ubuntu 24.04.2 LTS and Linux 6.8.0-57-generic. We use Occlum containers based on version 0.31.0-rc-ubuntu22.04. The configured EPC size is 64\;GB. We report the average of 5 runs.}. As expected, the TEE-enabled backend performs comparably to the baseline. However, the FHE backend exhibits significant overhead.
We will evaluate the impact of the different backends (execution time and scalability) on all the elementary operations as part of future work.

\section{Related Work}
\label{sec:rw}

Previous work have already explored DSL for secure computation. \cite{alchemy} proposes a DSL for FHE, \cite{hastee} proposes a DSL for computation in Intel SGX. Although these DSLs simplify the use of secure computing technologies, they focus on a single technology. Thus, they do not allow the use of a single code base in a hybrid environment.

\cite{google-tfhe-transpiler} proposes a FHE transpiler that allows to transpile a C++ code to TFHE circuits. \cite{Concrete} proposes a TFHE transpiler to transpile a Python function. By doing this, they allow developers to use an existing codebase on top of FHE. Howeover, the code must still be adapted to conform to the limitations imposed by the use of FHE. Furthermore, these tools are not designed for integration into a data processing workflow within a hybrid environment.

\begin{acks}
This work was supported by the Swiss National Science Foundation under project P4: Practical Privacy-Preserving Processing (no. 215216).
\end{acks}

\bibliographystyle{ACM-Reference-Format}
\bibliography{references}

\end{document}